\documentclass[subabstract]{aa}
\usepackage{amsmath,graphicx,amssymb}
\usepackage{longtable}
\usepackage{natbib}
\bibpunct{(}{)}{,}{a}{}{}

\newlength\staretab

\begin{document}

\title{THE MAGNITUDE-DIAMETER RELATION OF GALAXIES}

\author{Sidney van den Bergh}

\institute{National Research Council of Canada, Dominion Astrophysical Observatory, Herzberg Institute of Astrophysics, 5071 W. Saanich Rd., Victoria, BC, V9E 2E7, Canada}
            
\date{Received / Accepted}

\abstract{}{We investigate the dependence of the luminosity-diameter relation of galaxies on the environment.} {This study is based on a comparison between the 80 galaxies 
in the Shapley-Ames Catalog that are located within a distance of 10
Mpc, and the luminosity-diameter relation for galaxies in great
clusters such as Virgo and Coma.}{ A relatively tight linear
correlation is observed between the absolute magnitudes and the
logarithms of the linear diameters of galaxies located within 10
Mpc. Surprisingly this observed power-law relationship appears to be
almost independent of environment and local mass density as defined by
Karachentsev \& Malakov. However, at a given luminosity, early-type
galaxies are (on average) slightly more compact than are objects of
a later type. Unexpectedly the present results appear to indicate that
the luminosity-diameter relation for the galaxies within 10 Mpc is
indistinguishable from what is  observed in the much denser Virgo
cluster.} { Galaxies appear to
form an almost one-dimensional family in parameter space. It remains a
mystery why the luminosity-diameter relation for galaxies is so
insensitive to environment.}{}


\maketitle

\titlerunning{The Magnitue-Diameter Relation in Galaxies}
\authorrunning{Sidney van den Bergh}

\section{Introduction}

Giuricin et al. (1988) first drew attention to the surprising
fact that the luminosity-diameter relation provides the tightest of
all correlations observed among the photometric parameters that can be
used to characterize galaxies.  Subsequently this conclusion was
strengthened and confirmed by Gavazzi et al. (1996) who found a strong
correlation between diameter and blue-light luminosity of galaxies,
and an even tighter correlation between H-band luminosity and disk
galaxy diameter.  Similarly tight relationships can be obtained by
comparing photometric and dynamical parameters describing galaxies
(Faber \& Jackson 1976, Tully \& Fisher 1977). According to present
ideas the formation of galaxies is a very messy and chaotic process
involving both multiple mergers and complicated feed-back
effects. Furthermore the history of the rate of star formation in
individual galaxies can differ in a bewildering variety of ways. From currently fashionable ideas on galaxy formation one would expect
environment, initial mass, angula momentum, central concentration,
star formation history and merger history to all affect the evolution
of disk galaxies (Dalcanton et al. 1997, Ma et al. 1998).  This raises
the question why parameters, such as galaxy luminosity and diameter,
should still be so closely correlated at the end of widely differing
evolutionary tracks in different environments. Possibly some
hints about the origin of the observed  power-law relation between the
luminosity and the diameter of galaxies might be obtained by searching
for variations of the luminosity-diameter relation as a function of
galaxy environment, which may itself be characterized in a variety of
differing ways. Girardi et al. (1991) have pointed out that ``The
tightness of this relationship suggests that possible environmental
effects should be detectable from the analysis of its shape in
different samples". Giuricin  et al. (1988) claim to have
detected ``appreciable differences in the galaxy-luminosity
relationships for different clusters''. On the other hand Girardi et
al. (1991) reach the opposite conclusion and find no significant
differences between the luminosity-diameter relations in a variety of
environments. Similarly Gavazzi et al. (1996) obtain broadly
similar luminosity-diameterrelations in differing environments. It is
the purpose of the present study to re-investigate this question by
comparing the luminosity-diameter relations in  three
particularly high quality data sets: (1) The Shapley-Ames galaxies
with D $<$ 10 Mpc (2) the disk galaxies in the Virgo cluster, and
(3) a number of other more distant Abell clusters. 

\section{Database}

 Physical information is most complete for the nearest and the
 brightest galaxies. Such galaxies therefore provide an excellent
 laboratory for study of the relationship between the luminosities and
 diameters of galaxies. A catalog of 451 galaxies, that are believed
 to be closer than 10 Mpc, has recently been published by Karachentsev
 et al. (2004). A listing of the brightest galaxies (which is complete
 to B $\sim$13) was given by Shapley \& Ames (1932) and has been
 updated and revised by Sandage \& Tammann (1981). {\it The Revised
 Shapley-Ames Catalog}, which contains 1276 galaxies, is particularly
 valuable because it provides complete and homogeneous information on
 the morphologies of all of the brightest galaxies based on uniform
 classifications that were made by expert morphologists on the basis
 of inspection of plates that were almost all obtained with large
 reflecting telescopes. The Karachentsev et al. and Revised
 Shapley-Ames catalogs have 79 galaxies in common, or 80 if the Milky
 Way System (which was omitted from the Shapley-Ames Catalog) is
 included. A listing of all of these galaxies is given in Table 1. For
 each object this table lists: (1) The galaxy name, (2) the major axis
 linear diameter $A_{25}$ taken from the compilation of Karachentsev
 et al. (2004). This diameter was corrected for galaxy inclination and
 Galactic extinction in the manner of the RC2 Catalog (de Vaucouleurs
 et al. 1976). The value of $A_{25}$ would (after scaling to a
 common distance scale) be essentially the same as the diameter
 $D_{0}$ of De Vaucouleurs et al. and twice the value R used by
 Gavazzi et al. (1996).  In their catalog Karachentseve et al. (2004)
 indicate which method was used to derive the distance of each
 individual galaxy within a global framework in which a Hubble
 parameter of 72 km $s^{-1}$ Mpc$^{-1}$ was adopted, (3) The
 integrated magnituds blue $M_{B}$ was taken from Karachentsev et
 al. (2004), (4) the adopted tidal index $\theta$ is based on the
 local mass density (Karachentsev \& Malakov 1999). In this relation
 the zero-point has been set in such a way that $\theta$ = 0 when the
 Keplerian cyclical period of a galaxy, with respect to its main
 disturber, equals the cosmic Hubble time 1/$H_{o}$ . Galaxies with
 $\theta < $ 0 may be considered as undisturbed (isolated) objects. A
 caviat is, of course, that the real orbital periods will be affected
 by the unknown distribution of dark matter. (5) The adopted
 morphological types are slightly simplified versions of the
 classifications given in Sandage \& Tammann (1981). Finally, (6)
 galaxies north of $\delta$ = -27$^o$ (which are visible on the {\it
 Palomar Sky Survey}) were, on the basis of visual inspection,
 assigned to clusters (C), groups (G) or the field (F) environments
 (van den Bergh 2007) plus unpublished data.  Galaxies were assigned
 to groups if they appeared to have between three and five non-dwarf
 companions, or to clusters if they had six or more non-dwarf
 companions. The utility of this classification is attested to by the
 strong correlation that is observed between the C, G and F
 assignments and the intrinsic U-B colors of galaxies (van den Bergh
 2007).

It might be argued that a volume-limited galaxy sample, that includes
many dwarf galaxies and reaches out to a fainter limiting isophote,
would be physically more meaningful than the present sample of
Shapley-Ames galaxies within 10 Mpc. However, a distinct disadvantage
of such a sample is that it would include many objects for which the
photometric and morphological data are incomplete, or of low
quality. Finally many dwarf galaxies would be excluded from the
sample because their surface brightness never attains the 25 mag
arcsec$^{-2}$ surface brightmass that defines the $A_{25}$ diameter.

The fact that the Shapley-Ames sample of nearby galaxies, and the
Binggeli et al.(1984) sample of Virgo galaxies, are both luminosity-limited
facilities a comparison between the luminosity-density relations in
these two very different environments.  By the same token the data
compiled by Gavazzi et al. are subsamples of the Zwicky et
al. (1961-1968) catalog, which comprises luminosity selected galaxies
that that have B $\leq$ 15.7, allowing one to compare galaxies in some
Abell clusters with comparable data for field galaxies in the ``Great Wall''.

 Yet another approach might have been to examine the relation between the luminosities of galaxies and their half-light radii. However, this has the disadvantage that the $M_{B}$ versus  log $r_{e}$ relation is distinctly non-linear (Binggeli et al. 1984), whereas (over a range of many hundreds in luminosity) the relation between $A_{25}$ diameter and galaxy luminosity is a simple power law. Finally one might argue that it would have been better to use the infrared, rather than the blue, luminosities of galaxies.  Such I-band magnitudes are expected to exhibit a closer correlation with galaxy mass than is the case for B-band magnitudes.  However, any dependence on environmental factors is more likely to show up in the B images which are much more strongly affected by star formation and dust.

\section{Discussion}

Figure 1A shows a plot of galaxy diameter $A_{25}$ in kpc versus
galaxy absolute magnitude $M_{B}$ for all of the galaxies listed in Table 1. This figure shows that, over a remarkably large range in luminosity, galaxies exhibit a reasonably tight power-law relation between luminosities and diameter. This result is both unexpected and puzzling. One might have expected the chaotic merger history of galaxies, and their widely differing histories of star formation, to have produced a broad spectrum of size-luminosity relationships.
Furthermore the sizes of some galaxies are likely to have been affected by feedback produced by active galactic nuclei during early phases of their evolution. Figure 1 shows that, after reducing the Girardi et al.
(1991) observations to the distance scale used in the present paper, the objects in Table 1 are well-represented by the relation\\

M$_{B}$ = -12.65 -5.7 log A$_{25}$,  \hspace*{2.5cm}(1)\\

~~~~~~~~~$\pm$0.1 ~~~$\pm$0.2     \\

{\flushleft that Girardi et al. (1991) derived for 177 disk
galaxies in the Virgo cluster.  It should be emphasized that the
errors quoted in Eqn (1) are probably underestimated because they do
not include systematic effects that might result from the bias that is
introduced by incompleteness at the faint end of the galaxy sample
that these authors used.}

The rms dispersion of the galaxies listed in Table 1 derived
from the regression line defined by Eqn. (1) is 0.8 mag. From a
similar study of 533 disk galaxies in eight relatively nearby rich
clusters and in the inter-cluster ``Great Wall'' Gavazzi et al. (1996)
find a slope of 6.6, which is marginally greater than the value of 5.7
$\pm$ 0.2 that Girardi et al. (1991) found for the Virgo
cluster. Gavazzi et al. argue that the power-law relationship between
the radii and the masses of galaxies suggest ``that to the first order
the process of galaxy evolution can be described with a single
parameter; the initial mass". It may be challenging to reconcile this
conclusion with the hierarchical merging scenario of the $\Lambda$CDM
Scenario in which the initial mass concept appears meaningless.  In a different
context Woo et al. (2008) have recently found that the Local Group
dwarfs basically define a one-parameter ``fundamental line" primarily
driven by steller mass.

 In Figure 1 distance errors will cause a galaxy to slide along a line
 with slope -5.0 (corresponding to constant surface brightness),which
 is close to the  observed slope of -5.7 in Eqn. (1). As a result
 even moderately large errors in distance estimates for individual
 galaxies will not add significantly to the observed dispersion of
 galaxies around the line defined by Eqn. (1). The most strongly
 deviant object (marked by a plus sign) is NGC 1569 (see panel \#336
 of Sandage \& Bedke 1994), which is unusually small for its
 luminosity. Van den Bergh (1960) classifies it as Ir pec III-IV? It
 appears likely that this object is too bright for its diameter
 because it underwent a violent star-burst in the relatively recent
 past  (Greggio et al. 1998).  It is noted in passing that
 Grocholski et al. (2008) have recently used HST observations to show
 that NGC 1569 is more distant than previously believed and it
 probably a member of the IC 342 group.  From the data in Table 1 it
 is also seen that NGC 221 (M32) is smaller than would be predicted by
 Eqn. (1). The reason for this is, no doubt (Faber 1973),
 that M32 has been stripped of its outer envelope by one or more
 strong tidal encounters with M31.

The largest galaxy within a sphere of radius 10 Mpc is NGC 5457
(M101). This object is almost twice as large as the two next largest
galaxies with D $<$ 10 Mpc: The Andromeda galaxy and NGC 4258
(M106). Peebles (2007) has drawn attention to the fact that M101 is
located near the edge of the Local Void, rather than in the dense
central region of the Local Super-cluster.

Figure 1B shows that field (F), group (G) and cluster (C) galaxies are
all distributed similarly relative to the line defined by Equation
1. This result is slightly puzzling because van den Bergh (2007) found
a very strong correlation between the morphological types of galaxies
and their assignment to F, G and C environments. One might perhaps
have expected cluster members (which are mainly of early morphological
types) to be smaller than field galaxies that mostly have late
morphological types.

Figure 1C shows that field galaxies, and objects that are bound on
timescales $< 1/H_{o}$, are distributed in the same fashion. Taken at
face value these results suggest that the position of an object on the
luminosity-diameter relation is essentially independent of the
environment in which it finds itself. This result is again somewhat
counterintuitive because one might have expected compact early-type
galaxies to occur predominantly among objects that are bound on a
Hubble time, whereas more extended late-type galaxies would perhaps
have been expected to predominate among unbound field galaxies.

Finally Figure 1D shows that early-type and late-type galaxies are
displaced relative to each other. Not unexpectedly the figure shows
that, at a given luminosity, early-type galaxies are more compact than
galaxies of late type. In other words, the luminosity-diameter
relation for early-type galaxies is slightly displaced towards smaller
diameters relative to that for late- type galaxies.  These small
deviations show that galaxies are almost, but not exactly, a one
parameter family.  In this connection it is of interest to note that
Gavazzi et al. (1996) find that the power law relationship between
galaxy luminosity and diameter is even tighter in H-band than it is in
the I-band.  This no doubt is to the fact that environmental factors,
that may influence galaxy formation and the presence of dust will, affect the
blue luminosity of a galaxy more that they will its infra-red
luminosity.

One of the main result of the present investigation is that the
relatively well-studied Virgo cluster exhibits a luminosity- diameter
relationship that is indistinguishable from that of the lower density
region containing the Shapley-Ames galaxies that are situated within a
distance of 10 Mpc. In their study of the Virgo cluster Girardi et
al. (1991) divided their Virgo sample into three sub-regions: An inner
shell within R $\leq$ 0.5 Mpc of the center of the Virgo cluster, an
intermediate shell with 0.5 $< R \leq$ 1.0 Mpc, and an outer shell
with R $>$ 1.0 Mpc. These authors found that these three subsamples
exhibited extremely similar luminosity-diameter relations. More
fragmentary data that Girardi et al. collected on the diameters of
other (mainly more distant) clusters do not appear to show clear-cut
differences between the luminosity-diameter relation for the Virgo
cluster and those for other dense clusters.

Girardi et al. (1991) have also compared the luminosity-diameter
relations of objects in Tully's (1988){\it  Nearby Galaxies Catalog}, which
lists galaxies out to a redshift of 3000 km $s^{-1}$ corresponding to
R $\simeq$ 42 Mpc. Within this largere volume these authors find no
significant differences between the luminosity-diameter relations for
galaxies that Tully assigns to (1) the field, (2) groups, and (3)
clusters. Girardi et al. therefore conclude that ''These results
showed that any possible environmental effect is not strong enough to
affect significantly the L-D relation for disk galaxies in the samples
used in the present paper.'' The results obtained from Tully's catalog
are therefore entirely consistent with those shown in Figure 1B which
exhibits a similar independence from environmental effects for
galaxies that van den Bergh (2007) assigned to clusters, groups and
the field from visual inspection of the prints of the {\it Palomar Sky
Survey}. Furthermore these results are also consistent with the data
plotted in Figure 1C, which show no obvious dependence of the
luminosity-diameter relation among nearby galaxies on the $\theta$
index of Karachentsev \& Malakov (1999).  Finally Gavazzi et
al. (1996) find no systematic difference between the
luminosity-diameter relations in dense Abell clusters and among the
more isolated individual galaxies in the ''Great Wall''. The fact that
Gavazzi et al. find a much tighter correlation between H-band
luminosity and radius than between B-band luminosity and galxy radius
confirms the suspicion that environmental effects are smaller at long
wavelengths.

\section{Summary}
 The present data on nearby Shapley-Ames galaxies strengthen and
 confirm the conclusion by Gavazzi et al. (1996) that galaxies can be
 regarded as an (almost) one-parameter family in parameter space.  It
 will be challenging to reconcile the conclusion that galaxies are
 essentially defined by their initial mass with the presently
 fashionable hierarchical merging scenario. Perhaps this conundrum
 could be resolved by assuming that most of the mergers that occur in
 the hierarchical merger scheme are with low-mass galaxies.  Such
 mergers would add little to the mass of the main body of the dominant
 galaxy, although they might make major contributions to the stellar
 populations in the halos of the merger products.

The Shapley-Ames galaxies within 10 Mpc show a relatively tight
correlation between luminosity and diameter. Surprisingly this
correlation appears to be almost independent of a galaxy's environment
that has been characterized in two different ways. On a larger scale
no significant difference is found between the local
luminosity-diameter relation and that observed in the Virgo
cluster. Even within the Virgo cluster itself the luminosity-diameter
relationship appears to be essentially the same in the dense cluster
core and in its lower density envelope.  By the same token Gavazzi et
al. (1996) find no evidence for any systematic difference between the
luminosity-diameter relation in relatively nearby Abell clusters and
in the lower density Great Wall region.  These results are surprising
because, as Gavazzi et al point out, hierarchical models of galaxy
formation predict a dependence of galaxy photometric parameters on
environmental conditions.

 I thank Igor Karachentsev for a discussion of the nature of the
 power-law relationship between galaxy size and luminosity and also
 for information on NGC 4236. Furthermore I am also indebted to Marisa
 Girardi for correspondence about the distance scale used in her paper
 and to Thomas Puzia for his comments on a draft of this paper  in
 which he suggested that the published magnitude of NGC 1569 might
 have been affected by a nearby foreground star.  I am also indebted
 to Brent Tully and to an anonymous referee for a number of wise
 comments that were deeply appreciated. Thanks are also due to Brenda
 Parrish and Jason Shrivell for technical assistance.

\onecolumn
\longtab{1}{
\begin{longtable}{lrrrll}
\caption{\label{table} Data on all Shapley-Ames galaxies within 10 Mpc}\\
\hline\hline 
Name & $A_{25}$  & $M_{B}$ & $\theta$ & Type & Gr\\
\hline
\endfirsthead
\caption{continued.}\\
\hline \hline 
Name & $A_{25}$  & $M_{B}$ & $\theta$ & Type & Gr\\
\hline
\endhead
\endfoot

N  55   &  12.05 &  -18.06 &   -0.4  &  Sc      &     ...\\
N 147   &  3.16  & -14.79  &  + 3.0  & dE5      &     G\\
N 185   &  2.51   &-14.76  &   +2.3  &  dE3     &      G\\
N 205   &   4.49 &  -16.15 &   +3.7  &  S0/E5   &      G\\
N 221   &   1.99 &  -15.96 &   +6.8  &  E2      &      G\\
N 224   &  35.87 &  -21.58 &   +4.6  &  Sb I-II &      G\\
N 247   &  20.66 &  -18.81 &   +1.3  &  Sc III-IV &    G\\
N 253   &  22.98 &  -21.37 &   +0.3  &  Sc        &    G\\
SMC     &   5.28 &  -16.35 &   +3.5  &  Im IV-V   &   ...\\
N 300   &  12.95 &  -17.92 &   -0.3  &  Sc III    &   ...\\
N 404   &   3.25 &  -16.49 &   -1.0  &  S0        &    G\\
N 598   &  16.09 &  -18.87 &   +2.0  &  Sc II-III &    G\\
N 625   &   5.96 &  -16.53 &   -0.4  &  Amorph    &   ...\\
N 628   &  22.54 &  -19.84 &    0.0  &  Sc I      &    F\\
N 672   &  13.19 &  -18.76 &   +3.8  &  SBc III   &    F\\
N 891   &  30.80 &  -20.59 &   -1.2  &  Sb        &    F\\
N 925   &  28.60 &  -19.88 &   -0.9  &  SBc II-III&    F\\
N1313   &  11.45 &  -19.02 &   -1.6  &  SBc III-IV&   ...\\
N1569   &   3.34 &  -18.17 &   -0.4  &  Sm IV     &    F\\
N1705   &   2.67 &  -15.86 &   -1.7  &  Amorph    &   ...\\
LMC     &   9.75 &  -17.93 &   +3.6  &  SBm III   &    G\\
N2366   &   5.71 &  -16.02 &   +1.0  &  SBm IV-V  &    G\\
N2403   &  19.43 &  -19.29 &    0.0  &  Sc III    &    G\\
N2537   &   3.49 &  -16.65 &   -1.4  &  Sc III    &    F\\
N2683   &  16.18 &  -20.43 &   +0.2  &  Sb        &    F\\
N2784   &  15.86 &  -19.72 &   +2.0  &  S0        &    F\\
N2787   &   7.13 &  -18.50 &   -1.9  &  SB0/a     &    F\\
N2903   &  28.94 &  -21.00 &   +1.8  &  Sc I-II   &    F\\
N2976   &   5.57 &  -17.10 &   +2.7  &  Sd III-IV &    G\\
N3031   &  26.85 &  -21.06 &   +2.2  &  Sb I-II   &    G\\
N3034   &  10.93 &  -19.63 &   +2.7  &  Amorph    &    G\\
N3109   &   5.81 &  -15.68 &   -0.1  &  Sm IV     &    F\\
N3077   &   6.14 &  -17.76 &   +1.9  &  Amorph    &    G\\
N3115   &  17.13 &  -20.82 &   +1.9  &  S0        &    F\\
N3274   &   3.50 &  -16.16 &   -0.3  &  S IV      &    G\\
N3344   &  14.16 &  -19.03 &   -1.5  &  SBbc I    &    F\\
N3351   &  20.53 &  -19.88 &   +0.8  &  SBb II    &    C\\
N3368   &  21.82 &  -20.42 &   +0.6  &  Sab II    &    C\\
N3377   &  15.27 &  -19.10  &  +0.8   & E6        &    C\\
N3379   &  17.46 &  -20.10  &  +1.0   & E0        &    C\\
N3384   &  15.91 &  -19.55  &  +1.2   & SB0       &    C\\
N3412   &   9.94 & -18.76  &  +1.9    &  SB0      &     C\\
N3489   &   9.60 &  -19.45 &   +1.5   & S0/Sa     &    C\\
N3593   &   9.00 &  -17.77 &   -1.7   & Sa        &    G\\
N3621   &  22.05 &  -19.81 &   -1.9   & Sc III    &   ...\\
N3627   &  23.98 &  -21.14 &   -0.7   & Sb II     &    G\\
N3738   &   3.52 &  -16.61 &   -1.0   & Sd III    &    C\\
N4144   &  12.89 &  -18.25 &   -0.9   & Scd III   &    C\\
N4190   &   1.73 &  -14.33 &    0.0   & Sm IV     &    C\\
N4214   &   7.05 &  -17.19 &   -0.7   & SBm III   &    C\\
N4236   &  23.58 &  -18.59 &   -0.4   & SBd IV    &    G\\
N4244   &  16.60 &  -18.60 &    0.0   & Scd       &    C\\
N4258   &  35.61 &  -21.25 &   -0.7   & Sb II     &    C\\
N4395   &  17.30 &  -17.85 &   +0.1   & Sd III-IV &    C\\
N4449   &   7.21 &  -18.27 &    0.0   & Sm IV     &    C\\
N4460   &   8.92 &  -17.89 &   -0.7   & Sbc       &    C\\
N4594   &  21.10 &  -21.90 &   +0.3   & Sab       &    C\\
N4605   &   7.41 &  -17.96 &   -1.1   & Sc III    &    G\\
N4736   &  14.80 &  -19.83 &   -0.5   & Sab       &    C\\
N4826   &  20.41 &  -20.90 &   -1.7   & Sab II    &    F\\
N4945   &  17.41 &  -20.51 &   +0.7   & Sc        &   ...\\
N5068   &  14.76 &  -19.04 &   -1.4   & SBc II-III&    G\\
N5102   &   7.20 &  -18.08 &   +0.7   & S0        &   ...\\
N5128   &  28.88 &  -20.77 &   +0.6   & S0+Spec   &   ...\\
N5204   &   6.17 &  -16.75 &   -1.1   & Sd IV     &    C\\
N5194   &  24.44 &  -21.34 &   +4.1   & Sbc I-II  &    G\\
N5195   &  11.83 &  -19.22 &   +5.0   & SB0       &    C\\
N5236   &  17.37 &  -20.43 &   +0.8   & SBc II    &   ...\\
N5253   &   5.04 &  -17.38 &   +0.5   & Amorph    &   ...\\
N5457   &  61.44 &  -21.23 &   +0.6   & Sc I      &    C\\
N5474   &   9.95 &  -17.74 &   +2.0   & Scd IV    &    C\\
N5585   &   8.92 &  -17.82 &   -0.8   & Sd IV     &    C\\
Milky Way & 25.0: &  -20.8  &   +2.5   & Sbc:      &    G\\
I4662   &   1.55 &  -15.13 &   -0.9   & Im III    &   ...\\
N6503   &   9.03 &  -18.08 &   -1.2   & Sc III    &    F\\
N6822   &   2.71 &  -15.22 &   +0.6   & Im IV-V   &    G\\
N6946   &  25.96 &  -20.86 &   +0.7   & Sc II     &    F\\
I5052   &   7.64 &  -18.23 &   -2.2   & Sd        &   ...\\
I5152   &   2.91 &  -15.67 &   -1.1   & Sdm IV-V  &   ...\\
N7793   &   9.96 &  -18.53 &   +0.1   & Sd IV     &   ...\\

\hline
\hline

\end{longtable}
}

\begin{figure*}[ht]
\centering \resizebox{1\hsize}{!}{\includegraphics{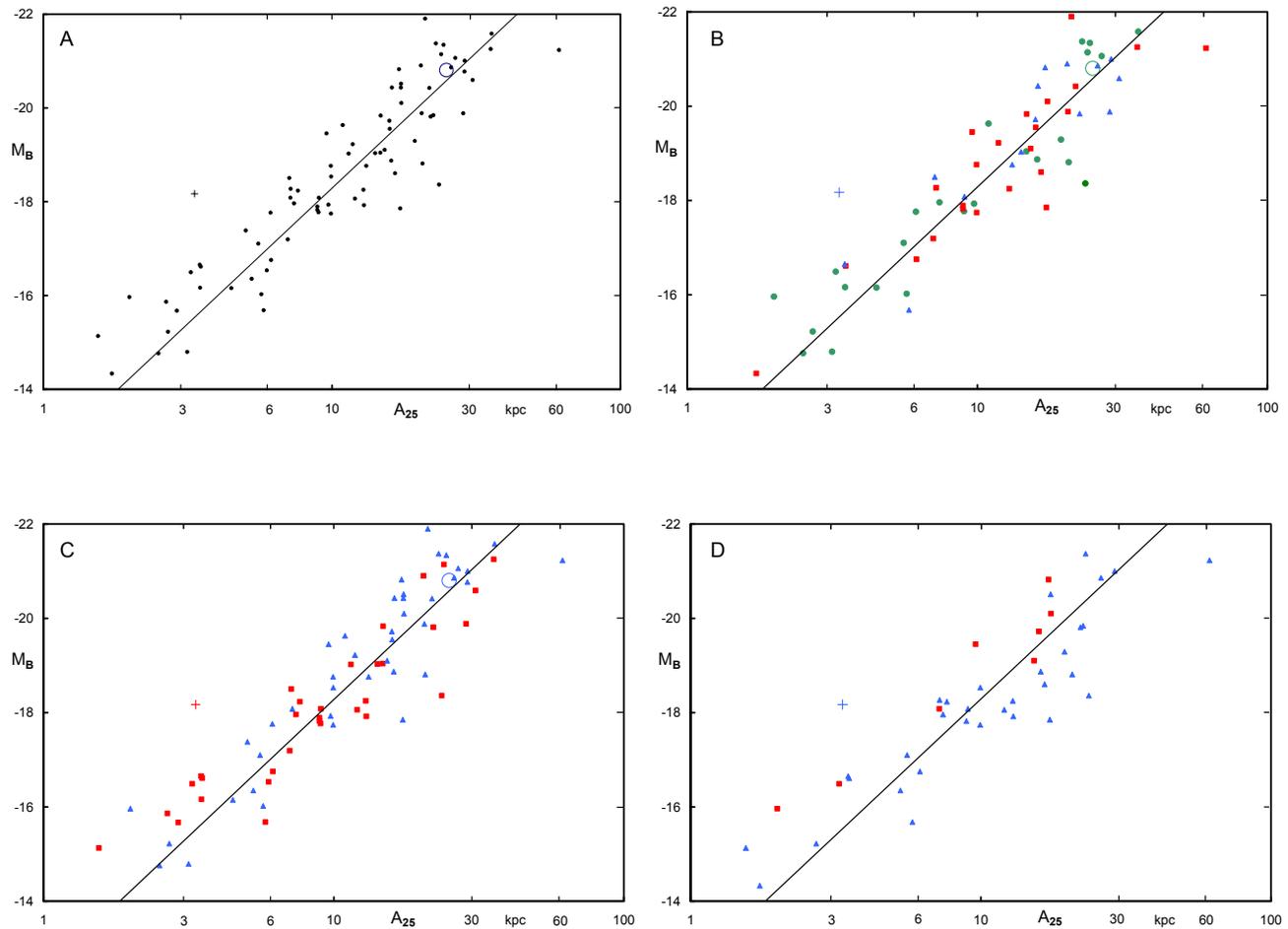}}
\caption{This plot shows the relationship between the luminosities and
the diameters of Shapley-Ames galaxies within 10 Mpc. Also shown as a
solid line is the luminosity-diameter relation for the Virgo cluster
according to Girardi et al. (1991).  Figure 1A shows that nearby galaxies and those in the Virgo cluster follow the same power-law relationship.  The deviant galaxy NGC 1569 is plotted as a plus sign.  The Milky Way system is shown as a small circle.  Figure 1B shows the Luminosity-Diameter relation for galaxies in clusters (red squares), in groups (green circles) and in the field (blue triangles).  The Milky Way System is shown as a green circle.  The figure shows a surprising absence of dependence of position in this diagram on environment.   Figure 1C shows the Luminosity-Diameter relation for galaxies with $\theta <  0$, which may be considered as undisturbed isolated objects (red squares).  Galaxies with $\theta > 0$, are plotted as blue triangles.  These objects have Keplerian cyclical periods smaller than the Hubble time $1/H_{0}$. The small blue circle represents the Milky Way System.  The plot shows no evidence for a clear-cut relationship between the Karachentseve \& Malakov tidal index and deviations from the mean luminosity-diameter relation.  Finally Figure 1D shows the luminosity-diameter relation for early-type (E+S0+S0/Sa) galaxies (red squares) and late type (Sc, Scd, Sd, Sm, Im) galaxies (blue triangles).  Not unexpectedly this figure shows that at, a given luminosity, early-type galaxies are (on average) slightly more compact than are objects of later morphological types.}
\end{figure*}





\end{document}